\documentclass[12pt,a4paper]{iopart}

\usepackage{iopams} 	
\usepackage{amssymb, amstext, amsopn} 
\usepackage{tabularx}
\usepackage{array}
\usepackage{multirow}
\usepackage{graphicx}
\usepackage{subfig}
\usepackage{hyperref}
\usepackage{epsfig}
\usepackage{epstopdf}

\DeclareMathOperator{\sinc}{sinc}
\def\tfrac{\case}

\def\be{\begin{equation}}
\def\ee{\end{equation}}

\begin{document}

\title{Mid-band gravitational wave detection with precision atomic sensors}

\author{Peter W. Graham}
\address{Stanford Institute for Theoretical Physics, Department of Physics, Stanford University, Stanford, CA 94305}

\author{Jason M. Hogan}
\address{Department of Physics, Stanford University, Stanford, CA 94305}
\ead{hogan@stanford.edu}

\author{Mark A. Kasevich}
\address{Department of Physics, Stanford University, Stanford, CA 94305}

\author{Surjeet Rajendran}
\address{Berkeley Center for Theoretical Physics, Department of Physics, University of California, Berkeley, CA 94720}

\author{Roger W. Romani}
\address{Kavli Institute for Particle Astrophysics and Cosmology}
\address{Department of Physics, Stanford University, Stanford, CA 94305.}

\author{For the MAGIS collaboration}

\begin{abstract}

We assess the science reach and technical feasibility of a satellite mission based on precision atomic sensors configured to detect gravitational radiation.  Conceptual advances in the past three years indicate that a two-satellite constellation with science payloads consisting of atomic sensors based on laser cooled atomic Sr can achieve scientifically interesting gravitational wave strain sensitivities in a frequency band between the LISA and LIGO detectors, roughly 30 mHz to 10 Hz.  The discovery potential of the proposed  instrument ranges from from observation of new astrophysical sources (e.g.~black hole and neutron star binaries) to searches for cosmological sources of stochastic gravitational radiation and searches for dark matter.

\end{abstract}

\maketitle

\section{Overview}

The recent first direct detections of gravitational waves by LIGO represent the beginning of a new era in astronomy \cite{PhysRevLett.116.061102,TheLIGOScientific:2016qqj,PhysRevLett.119.161101}.  Gravitational wave astronomy can provide information about astrophysical systems and cosmology that is difficult or impossible to acquire by other methods.  Just like with electromagnetic waves, there is a wide spectrum of gravitational wave signals that must be explored to fully take advantage of this new source of information about the universe.  LIGO and other ground-based laser interferometer designs are sensitive to gravitational waves between about 10 Hz and 1 kHz, but are severely limited at lower frequencies due to coupling to seismic noise \cite{TheLIGOScientific:2016agk,smith2009path}.  At much lower frequencies, the future space-based laser interferometer LISA is targeted at the 1 mHz - 50 mHz range \cite{Bender1998}.  Thus there is an opportunity for atomic sensors to play a key role in filling the gap between the LISA and LIGO detectors, which here we refer to as ``mid-band'' (approximately 30 mHz to 10 Hz).  The discovery potential of such instrumentation appears exciting, ranging from observation of new astrophysical sources to searches for cosmological sources coming from the early universe, as well as searches for ultralight dark matter.

To access this frequency range we propose a Mid-band Atomic Gravitational Wave Interferometric Sensor (MAGIS).  In the proposed detector concept, gravitational radiation is sensed through precise measurement of the light flight time between two distantly separated (atomic) inertial references, each in a satellite in Medium Earth orbit (MEO).  Ensembles of ultra-cold atomic Sr atoms at each location serve as precise atomic clocks and as precise inertial references using recently developed atom inteferometry and optical atomic clock methods based on optical interrogation of narrow-line (clock) transitions in Sr.  Light flight time is measured by comparing the phase of laser beams propagating between the two satellites with the phase of lasers referenced to the Sr optical transitions, accomplished using a phase meter and differential wave-front sensor similar to that to be demonstrated in GRACE-FO \cite{schutze_measuring_2016} and proposed for LISA \cite{danzman_lisa_2011,klipstein_2010}.  Since the atoms serve as precision laser frequency references, only two satellites operating along a single line-of-sight are required to sense gravitational waves.  In MEO orbits, the measurement baseline re-orients on a rapid time scale compared to the observation duration of the gravitational wave signals from many anticipated sources.  This allows efficient determination of sky position and polarization information.  The relatively short measurement baseline allows for excellent strain sensitivity in the 0.03 Hz to 3 Hz observation band, intermediate between the LISA and LIGO antenna responses, and suited to gravitational wave astronomy, cosmology and dark matter searches, as described below.

\section{Science objectives}

The gravitational wave (GW) spectrum between roughly 0.03 Hz and 3 Hz appears to be scientifically rich \cite{mandel2017astrophysical}. For example the lower end of this range (near 0.03 Hz) would allow observation of white dwarf binary mergers, while the higher frequency region (around 1 Hz) is potentially a valuable region for searching for more speculative cosmological sources such as inflation and reheating.  There are also sources that may be observed both in this band and in the LIGO or LISA bands.  For example, sources such as black hole or neutron star binaries may be observed in this band, and then observed later by LIGO once they pass into the higher frequencies.  Such joint observation would be a powerful new source of information, giving for example a prediction of the time and location of a merger event in LIGO which would allow narrow-field high sensitivity optical, x-ray, gamma ray, and other telescopes to repoint to observe the prompt emission during the actual merger event.  Since the sources generally live a long time in this mid-frequency band, they can be localized on the sky even by a single-baseline detector since the detector will change orientation and position significantly during the time spent observing a single source \cite{Graham:2017lmg}.

The detector proposed here could operate either in broadband \cite{hogan_atom_2015} or resonant mode \cite{graham2016resonant}.  In the resonant mode the resonance frequency can be chosen anywhere in the range roughly 0.03 Hz to 3 Hz.  Switching between the different modes or different resonant frequencies can be done rapidly by simply changing the sequence of laser pulses used, without changing  hardware or satellite configuration.  The question of the optimal observing strategy for such a resonant detector is non-trivial and requires additional study beyond the scope of this work. For the purpose of the science case outlined here, we will assume a rather simple observation strategy, and give numbers of sources observable with that strategy (with SNR of 3).  However there may well be a better strategy: the strategy chosen depends to some extent on the type of source one wants to search for, so it could be reoptimized for different sources.  The example choice of observing strategy we consider here, motivated by searching for the black hole binaries visible in LIGO as well as merging white dwarf binaries, is to sit at a frequency around 0.05 Hz (with a resonant $Q \sim 10$) in `discovery mode', waiting for a source to enter the band.  Once a source is discovered it can be tracked for longer by sweeping the detector frequency up to follow the source.  This will allow a significant improvement in SNR beyond discovery, allowing us to improve the measurement of source parameters and in particular to improve the angular localization on the sky.

\begin{figure}
\footnotesize
\begin{center}
\includegraphics[width=0.9\textwidth]{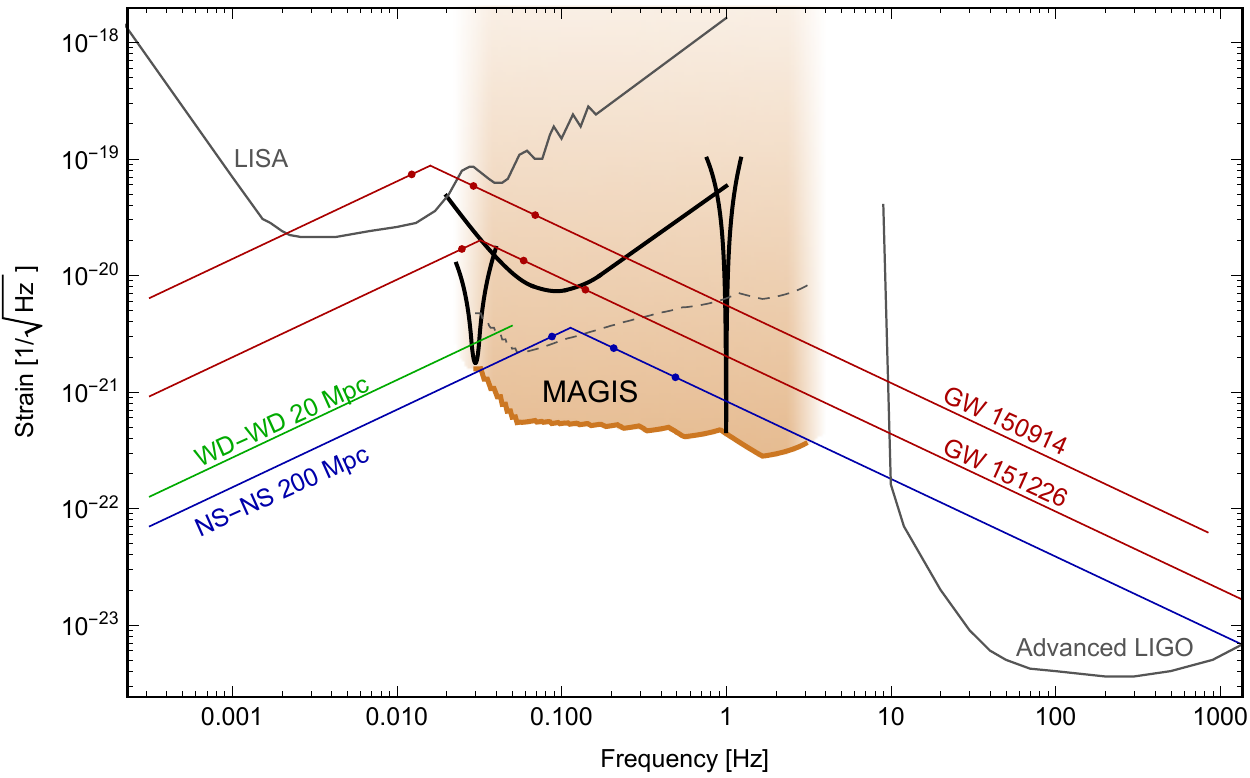}
\caption{
MAGIS gravitational wave sensitivity.  Two example resonant sensitivity curves (at arbitrarily chosen frequencies) and the broadband mode sensitivity are shown in solid, thick black.  The envelope of the possible resonant curves is shown by the lower brown boundary/line (this is the appropriate curve to compare to for discovery of longer-lived sources such as WD and NS binaries).  The dashed gray curve is the appropriate, approximate sensitivity curve for the discoverability (though not the ultimate measurement SNR) of shorter-lived sources such as the LIGO BH binaries.  The LISA strain curve is shown for reference \cite{Bender1998,larson2000sensitivity}. The Advanced LIGO curve is the design sensitivity \cite{smith2009path} (not current sensitivity \cite{TheLIGOScientific:2016agk}).  Two black hole merger events observed by LIGO are shown in red.  A WD-WD  binary at 20 Mpc (with masses $0.5 M_\odot - 0.5 M_\odot$) is shown in  green.  A NS-NS binary at 200 Mpc (with masses $1.4 M_\odot - 1.4 M_\odot$) is shown in blue.  Note that at frequencies in LIGO's band near the merger, the post-Newtonian formula we have used to draw these source curves breaks down.  The dots on the GW150914, GW151226 and NS-NS 200 Mpc curves indicate remaining lifetimes of 10 yrs, 1 yr and 0.1 yrs (reading left to right).
Degree scale sky localization appears feasible for in-spiraling sources.}
\label{fig:sensitivity}
\end{center}
\end{figure}

MAGIS' ability to provide advanced space-time localization (small angular error ellipses and precise chirp dates) of NS-NS, NS-BH and typical stellar BH-BH mergers is a key feature of the mission, since it enables simultaneous electromagnetic telescope observations.  A preliminary investigation of MAGIS' ability to localize the space-time position of a population of inspiral events has been made \cite{Graham:2017lmg}, but more detailed studies will be needed to explore the full set of capabilities of the proposed instrument and the science that results from this. Degree and $\sim 10^5$\,s -scale localization can certainly be achieved for sufficiently bright sources -- this allows wide-field contemporaneous E-M study w/ e.g. LSST, WFIRST and SKA. But with appropriate scheduling of resonant observing, some sources may also be localized to the 10-20 arcmin scale  \cite{Graham:2017lmg} which brings the full suite of modern astrophysical facilities into play, from the VLA through 10m-class telescopes to Chanda-type X-ray studies. This promises rich probes of the source physics, with a few examples given below.

\subsection{White Dwarf Binaries}

A double white dwarf (WD-WD) binary with secondary mass $M_2 M_\odot$ ends its inspiral (reaches Roche lobe contact) at $\sim 0.06 M_2$ Hz. Thus while a typical WD-WD radiates in the LISA band, the most interesting double WD, the candidate Type Ia progenitors with $M_1+M_2>1.4M_\odot$ reach the mid-frequency band. Using the rates in \cite{Badenes:2012} we expect a few such systems in the Milky Way, where high S/N allows us to measure departures from pure GR evolution, probing dissipative effects. With $\sim 150$ WD-WD Ia progenitors visible to Virgo group, we expect that a low-frequency MAGIS scan can measure the population of double degenerate (WD-WD) Ia progenitors. Lower mass WD-WD binaries are more common and a low frequency MAGIS search could detect several thousand sources, with $\sim 1$/y passing from the GR-dominated regime. The relevant sensitivity is shown by the brown line in Figure \ref{fig:sensitivity}, where narrow band sensitivity increases greatly toward 0.03 Hz. Measuring the high frequency population of the WD-WD spectrum thus tests the efficacy of this channel in producing Ia supernovae.

Rapidly spinning WD in close binaries provide a second class of WD sources in the MAGIS band. For example HD 49798 ($f_{GW} = 0.15$Hz) and AE Aq ($f_{GW}=0.06$Hz) offer two nearby ($d<$kpc) targets. These are measurable with S/N$\sim$10 in 30d resonant searches (after orbital de-modulation) if the WD quadrupole moment is $Q\sim 10^{47} {\rm g\, cm^2}$.  
Measuring the detailed waveforms (and quadrupole moments of spinning WD) probes the interior structure and merger dynamics.

\subsection{Binary Black Holes}

The pioneering discoveries at Advanced LIGO show that the universe is rich in binary black hole mergers \cite{TheLIGOScientific:2016pea} and encourages deeper exploration of the black hole population. When operated in the discovery mode at a frequency $\sim$ 0.05 Hz, MAGIS easily detects large chirp mass binaries such as the $\sim 26 M_{\odot}$ GW150914 event. This event is detectable out to $\sim 2$ Gpc with SNR $\gtrapprox 3$ (see Figure \ref{fig:sensitivity}), in  `discovery mode.'  This can be seen by comparing the source curve with the dashed gray line (since at these frequencies this source is relatively short-lived). More frequently MAGIS should also detect black hole binaries with more typical $8 M_\odot + 5 M_\odot$ ($M_{chirp} =5.5 M_\odot$) at frequencies $\sim 0.03-0.06$\,Hz. These would be discovered several years before the merger event, giving MAGIS the opportunity to pursue deeper studies using resonant sensitivity to predict the merger date in the Advanced LIGO band. For the louder sources, these predictions will include high quality pre-merger parameters, and degree-scale localization, allowing deep {\it targeted} radio/optical/X-ray searches for (presently unexpected) E-M signals during the merger events. Thus the MAGIS-AdvLIGO combination offers opportunities for powerful probes of the merger physics.

With a long lifetime in the MAGIS band, BH binaries may be intensively
studied to provide precise positions and pre-merger parameters. For higher
SNR events, study over many MAGIS orbits allows us to measure gravitational
wave polarization and the orientation of the binary orbit. Together with
AdvLIGO measurements of the chirp, this improves constraints on the
initial spins. If precessional or eccentricity effects are seen in the
pre-chirp signal, additional constraints may be extracted. Initial spins and
eccentricity are important clues to the BHs' origin and can be used to probe
a variety of astrophysics and fundamental physics scenarios \cite{Nishizawa:2016eza}.

For source localization, consider an event like GW150914. This event can be detected by MAGIS with a SNR $\sim 10$ when operated in the discovery mode. Subsequent to discovery, the signal can be resonantly followed, permitting a total SNR $\sim 50$ which should allow degree scale angular localization.
Further studies are necessary to optimize such measurement strategies.

\subsection{Neutron Star Binaries}

The recent detection of an NS-NS merger by LIGO and VIRGO and the subsequent observation by many EM telescopes (see e.g.~\cite{PhysRevLett.119.161101, GBM:2017lvd, Monitor:2017mdv}) has proven that neutron star binaries will be an exciting source for gravitational wave telescopes.  We have learned a lot from this first observation, and will continue to learn more as more sources are observed.  However there is still a great deal of information to extract from these events, for example better cosmological measurements  of the Hubble constant \cite{Abbott:2017xzu}.  It would be very useful to have warning and angular localization of the object prior to merger, as it allows one to target the prompt (ms to hour scale) counterpart emission with our most sensitive E-M facilities, from X-ray space telescopes to large aperture gound-based telescopes with photometric, spectroscopic and polarization sensitivity.  Such localization is difficult to achieve in the higher frequency band of LIGO and VIRGO.

As can be seen from  Fig.~\ref{fig:sensitivity}, the mid-frequency band is an ideal range for discovering neutron star binaries as well.
MAGIS could detect neutron star (NS) binaries out to roughly 300 Mpc (at SNR 3).
Note that to find the SNR of these NS detections at the lower frequencies of `discovery mode' one should compare the source curve in Figure \ref{fig:sensitivity} to the brown (not dashed gray) line.
This would allow a valuable connection with Advanced LIGO which can observe NS-NS binary mergers out to about 200 Mpc at their design sensitivity.  LIGO expects about 40 merger events within 200 Mpc per year.

The best strategy for discovering a population suitable for joint MAGIS AdvLIGO band study is likely an early-mission deep resonant survey at $f\sim 0.15-0.3$\,Hz where the detected sources will sweep into the AdvLIGO band in 1-5y. A subset of such sources suitable for intense study can be identified and follow-on resonant observations can be scheduled into a mission plan that optimizes science from these key sources, while exploring other mission objectives. The evaluation of such a mission schedule/observing strategy is beyond the scope of this work and is left for future study.

Optimized strategies need to be identified for using MAGIS broadband and resonant
observation time to best identify a sample of NS-NS and NS-BH in-spirals and provide
the maximum precision on the position and in-spiral date as alerts to AdvLIGO
and to electromagnetic facilities. Scheduling (months-years in advance)
simultaneous observations (optical, X-ray, gamma ray, etc.) for a meaningful
sample of chirp events will be a game changer. The full astrophysics community
can then participate in the study of the expected (but poorly predicted) E-M
signals from precursor through ring-down phases, teasing out unique information on the
neutron star equation of state, the relativistic MHD of the merger event and
the origin of the r-process elements. Without MAGIS, we have only shallow, sky
survey-type E-M observations of the prompt signal, with other telescopes observing the object only a long time period after merger, severely restricting the possible
science investigations.

\subsection{Intermediate Mass Black Holes}

If intermediate mass black holes (IMBH) exist
MAGIS should be able to discover inspirals of stellar mass compact objects (neutron stars and black holes) into $\sim \, 10^3 \, M_{\odot}$ primaries throughout much of the Hubble volume, with SNR $\sim 5$ when operating in the discovery mode around $\sim 0.05$ Hz. Little is known about this potential population (possible products of first generation stars and plausible seeds of the supermassive black holes powering AGN), but at MAGIS sensitivities, estimates suggest $\sim 1$ event per year for both black hole and neutron star inspirals \cite{Mandel:2007hi}. These large mass-ratio events offer unique opportunity to probe black hole spacetime \cite{Cardoso:2016ryw}. Of course, if binary black holes in the $10^3-10^5 M_{\odot}$ range exist they would be even louder and detectable to $z\sim 10$. Rate estimates are as high as $\mathcal{O}\left(10 - 100\right)$ per year \cite{Erickcek:2006xc}.  An interesting IMBH candidate has also recently been discovered \cite{2017NatAs...1..709O}.

\subsection{Cosmological Sources}
One conclusion of the estimates above is that conventional astrophysical sources, while representing extreme members of their object classes providing exceptional science yield, are relatively limited in number.  This is especially true in the upper half of the MAGIS frequency band. Above 0.05\,Hz we expect only a few Galactic rotating WD and WD-WD binaries, a few 10s of local group WD systems, and a few hundred NS-NS, NS-BH and BH-BH binaries passing through the MAGIS band.  At frequencies above the confusion-limiting irreducible LISA background of Galactic WD-WD binaries and below the 0.1-3\,kHz AdvLIGO range rich with NS and BH chirp events, MAGIS sources will be relatively limited and, with multi-year evolution times, nearly monochromatic. This exciting, but resolvable population of foreground sources leaves a large fraction of the MAGIS band open to searches for background cosmological sources and other exotica \cite{graham2016resonant}.

It has long been recognized that a detector around 1 Hz would be optimally suited to search for cosmological sources of gravitational waves such as inflation.  This is an extremely important signal to search for and is worth significant effort as gravitational waves offer possibly the only way to directly observe the universe before last scattering.  The only known ways to directly observe these signals from inflation are the CMB and direct gravitational wave detection in this frequency band.  The 1 Hz band probes a very different part of the inflationary epoch than the CMB, and a very different part of the inflaton potential.  Ideally, a detection in both the CMB and a direct gravitational wave detector would provide a powerful test of the scale-invariance of the inflationary spectrum over roughly 18 orders of magnitude in frequency, testing the heart of the inflation mechanism.  Further, such a long lever-arm would allow precise measurement of inflationary parameters which may not be testable any other way. MAGIS does not have the sensitivity to detect gravitational waves produced by typical slow roll inflation models that obey constraints imposed by the CMB. However,  unlike the CMB,  MAGIS probes the inflationary potential towards the end of inflation. This period is observationally unconstrained and may feature enhanced gravitational wave production. For example, MAGIS might discover models such as axion inflation  \cite{Cook:2011hg}.

MAGIS will also be sensitive to GWs produced by cosmic strings with tension $G\mu \sim 10^{-16}$.  First order phase transitions in the early universe at temperatures $\sim$ 100 TeV may also produce gravitational waves in the MAGIS frequency band. In light of the absence of new physics at the Large Hadron Collider that can explain the nature of electroweak symmetry breaking, the 100 TeV scale has emerged as a new target for particle physics \cite{Arkani-Hamed:2015vfh}. This scale also naturally emerges in frameworks such as the relaxion where the hierarchy problem is solved through cosmological evolution  \cite{Graham:2015cka}. If gravitational waves from the 100 TeV universe are discovered, it would open a dramatic path forward in particle physics.

\subsection{Dark Matter Direct Detection}

The identification of the properties of dark matter is a major scientific goal that has the potential to unveil new paradigms in astrophysics, cosmology and particle physics. Observational constraints allow the existence of a number of well motivated dark matter candidates. Ultra-light scalars are a particularly interesting class of dark matter.  They  emerge naturally in ultra-violet theories of particle physics such as string theory \cite{Arvanitaki:2009fg}, could explain facets of structure formation \cite{Hui:2016ltb} and may even play a central role in solving the hierarchy problem \cite{Graham:2015cka}. 

Single baseline gravitational wave detectors have enhanced sensitivity to ultra-light scalar dark matter  \cite{Arvanitaki:2016fyj}. These dark matter candidates can cause temporal oscillations in fundamental constants with a frequency set  by the dark matter mass, and amplitude determined by the local dark matter density, resulting in a modulation of atomic transition energies. This is a narrow signal in frequency space, with a maximum width $10^{-6}$ of the dark matter mass, set by virial velocity of the dark matter in the galaxy.  This spectral characteristic would distinguish it from backgrounds as well as gravitational wave signals. Further, unlike the gravitational wave signal which will change as the satellites move in orbit, these dark matter signals will be constant over timescales $\sim$ 1 - 10 years for signals in the frequency band 0.1 Hz - 1 Hz, enabling additional discrimination. This phenomenology is expected in dark matter candidates such as the relaxion that solve the hierarchy problem \cite{Graham:2015cka}. 

This signal is optimally measured by comparing two spatially separated atom-clock-like interferometers referenced by a common laser. Such a detector can improve on current searches for electron-mass or electric-charge modulus dark matter by up to 10 orders of magnitude in coupling, in a frequency band complementary to that of other proposals \cite{Arvanitaki:2016fyj}. MAGIS is ideally suited to perform this measurement. 

Ultralight dark matter can also potentially be searched for in other ways using this instrument, though this can require different modes of operation.  Axion-like dark matter and dark photons can be searched for through  the precession of nuclear spins they cause \cite{Graham:2017ivz}.  Scalar and vector dark matter can also be searched for through time-varying equivalence principle violations \cite{Graham:2015ifn} or possibly, depending on orbital configuration, through variations in the earth's gravitational field \cite{Geraci:2016fva}.

\section{Instrument description}

\subsection{Measurement Strategy}\index{Measurement Strategy}

\begin{figure}[!htbp]
    \centering
     \includegraphics[width=0.7\textwidth]{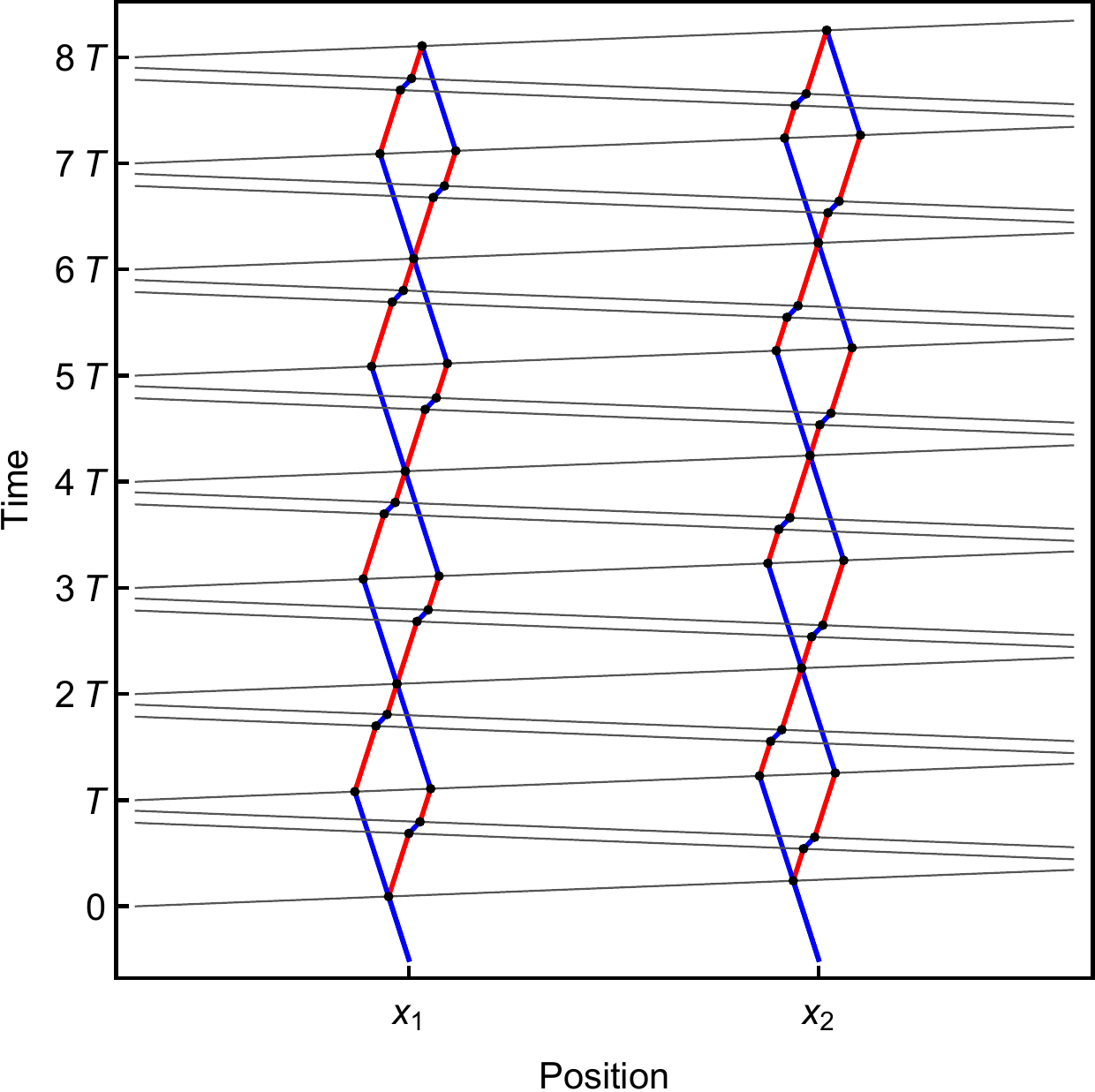}
    \caption{A space-time diagram of an example atom
interferometer detector sequence for the proposed MAGIS detector.  The detector consists of two atom interferometers based on single-photon transitions, one at position $x_1$ and the other at $x_2 = x_1+L$, where $L$ is the baseline distance.  The trajectories of the atoms are shown in blue for the ground state and red for the excited state. Pulses of light (thin gray lines) are sent back and forth from each end of the baseline and interact with the atoms (interactions shown as black dots), transferring momentum to the atoms and changing their internal state.  Whether or not an interaction occurs is controlled by matching the frequency of the light pulses to the Doppler shift of the atoms.  The sequence shown consists of two single-photon transitions for each atom optic ($n=2$, $2\hbar k$ momentum transferred) and a resonant enhancement of $Q=4$ (four diamonds).  The amount of resonant enhancement can be varied as needed by changing the pulse sequence.}
    \label{fig:resonantSpaceTimeDiagram}
\end{figure}

A space-time diagram of the proposed atom-based gravitational wave (GW) antenna is shown in Fig.~\ref{fig:resonantSpaceTimeDiagram}.  Dilute clouds of ultracold atoms at either end of the baseline act as inertial test masses, and laser light propagates between the atoms.  To implement atom interferometry, the lasers from each end of the baseline are briefly pulsed a number of times during each measurement cycle.  The paths of these light pulses appear as gray lines in Fig. \ref{fig:resonantSpaceTimeDiagram}.  The lasers are separated by a large distance $L$, with atom interferometers operated near them.  Interaction with a light pulse transfers momentum $\hbar k$ to the atom and toggles the atom between the ground and excited states.  As a result, the light pulses act as beam splitters and mirrors for the atom de Broglie waves, dividing them into a quantum superposition of two paths and eventually recombining them. Similar to an atomic clock, the phase shift recorded by each atom interferometer depends on the time spent in the excited state, which here is directly tied to the light travel time ($L/c$) across the baseline. GWs can be detected because they modulate the light travel time.  In Fig.~\ref{fig:resonantSpaceTimeDiagram}, an example resonant sequence is shown consisting of multiple diamond-shaped loops with a duration tuned to the GW period of interest \cite{graham2016resonant}.

A single interferometer of the type described above will be sensitive to laser noise [e.g., the interferometer at position $x_1$ in \ref{fig:resonantSpaceTimeDiagram}], but this is substantially suppressed by the differential measurement between the two interferometers at $x_1$ and $x_2$ \cite{Graham2013a}.  The two widely separated atom interferometers are run using common laser beams and the differential phase shift is measured.  The differential signal is largely immune to laser noise.  This cancellation of laser noise is enabled by the use of single-photon atomic transitions for the atom optics \cite{Graham2013a}.

\begin{figure}
    \centering
    \includegraphics[width=\textwidth]{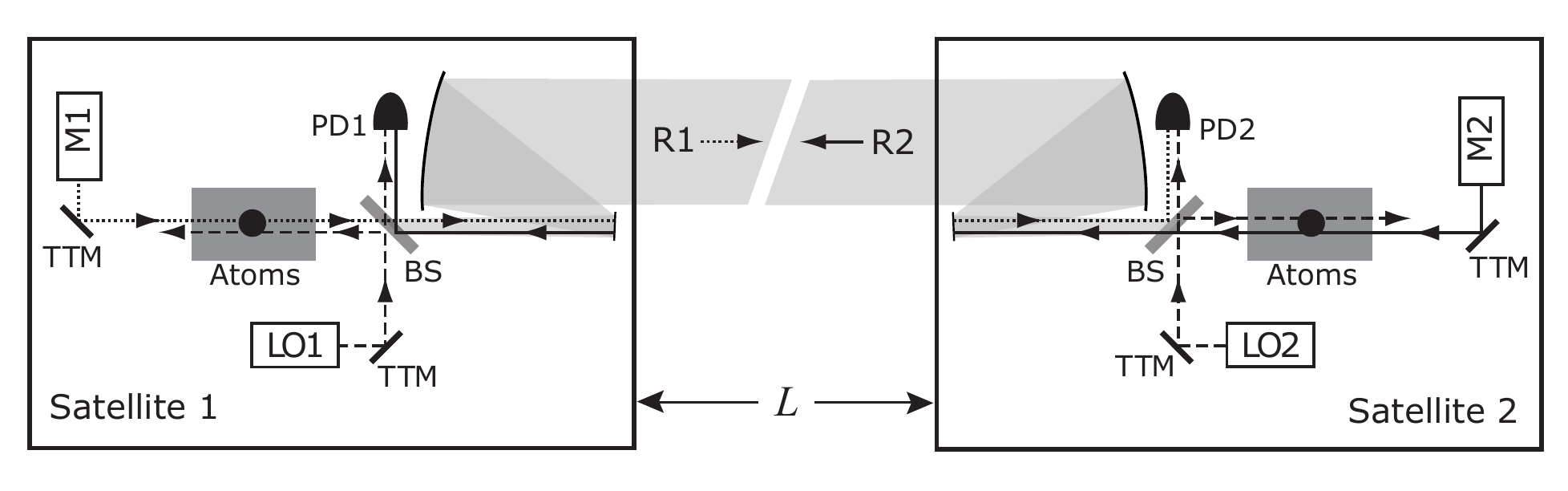}
    \caption{Schematic of the proposed design. M1 and M2 are the master lasers, with beams depicted as dotted and solid lines, respectively. The reference beams propagating between the satellites are denoted R1 (dotted) and R2 (solid) and originate from telescopes each each satellite.  LO1 and LO2 are local oscillator lasers (dashed beam lines) that are phase locked to the incoming reference laser beams (R2 and R1, respectively). PD1 (PD2) is a photodetector used to measure the heterodyne beatnote between the incoming reference beam R2 (R1) and the local oscillator laser LO1 (LO2) in order to provide feedback for the laser link. BS is a (non-polarizing) beam splitter where the heterodyne beatnote is formed. Tip-tilt mirrors (TTM) allow for fine control of the pointing direction of each laser. All adjacent parallel beams are nominally overlapped, but for clarity they are shown here with a small offset.}
    \label{Fig:apparatus}
\end{figure}

In the proposed implementation, intense local lasers are used to operate the atom interferometers at each end of the baseline.  To connect these otherwise independent local lasers, reference laser beams are transmitted between the two spacecraft, and the phases of the local lasers are locked/monitored with respect to the incoming wavefronts of these reference lasers. A conceptual schematic is shown in Fig. \ref{Fig:apparatus}.  A detailed description is available in Refs. \cite{hogan_atom_2015,graham2016resonant}.

In addition to a photodetector for measuring the phase difference between the two beams, a quadrant detector (or camera) is used to characterize the spatial interference pattern.  This allows the pointing direction and spatial mode of the two lasers to be well matched using appropriate feedback.  Feedback applied to the tip-tilt mirror (show as TTM in Fig.~\ref{Fig:apparatus} before the BS) can then be used to control the angle of the LO laser. Similarly, the angle of the master laser itself can be controlled by comparing it to LO laser direction and using another tip-tilt mirror.

The pulse sequences used to drive the interferometer can be changed to resonantly enhance the sensitivity of the detector \cite{graham2016resonant}.  The interferometer can be run in a resonant mode by using the pulse sequence $\pi/2 - \pi -  \dots - \pi - \pi/2$ with $Q$ $\pi$ pulses instead of the standard, broadband $\pi/2 - \pi - \pi/2$ pulse sequence.  These pulses are equally spaced in time by $T \lessapprox  T_{\text{max}}/Q$. Gravitational waves with frequencies $\omega = \pi/T$ oscillate by half a cycle between the pulses (e.g., from crest to trough). Unlike the broadband case, the phase differences caused by subsequent oscillations continually add in the resonant sequence, since the series of $\pi$ pulses periodically swap the arms of the interferometer.  To take advantage of large momentum transfer (LMT) phase enhancement, each atom optics `pulse' itself actually consists of $n$ closely spaced $\pi$-pulses which increase the separation of the interferometer arms \cite{kovachy2015quantum}.

\subsection{Instrument sensitivity model}\index{Instrument sensitivity model}\label{Sec:SensitivityModel}

Gravitational waves modify the light travel time across the baseline, modulating the time spent in the excited state by atoms on each side of the baseline and resulting in a differential phase shift between the two atom interferometers.  The phase response of the detector has the form $\Delta\Phi_\text{grad}(t_0)=\Delta\phi \cos{(\omega t_0+\phi_0)}$, where $\omega t_0+\phi_0$ is the phase of the GW at time $t_0$ at the start of the pulse sequence.  The amplitude of the detector response is \cite{graham2016resonant}
\begin{equation}
\Delta\phi= k_\text{eff} h L \frac{\sin\!{(\omega Q T)} }{\cos\!{(\omega T/2)}} \sinc{\big(\tfrac{\omega n L}{2 c}\big)}  \sin\!{\big(\tfrac{\omega T}{2}\!-\!\tfrac{\omega(n-1)L}{2c}\big)}\label{Eq:PhaseShift}
\end{equation}
Here $\hbar k_\text{eff}$ is the momentum transferred to the atom during the LMT beamsplitters and mirrors, where $k_\text{eff}\equiv n \omega_A/c$ for an $n$-pulse LMT sequence using an optical transition with atomic energy level spacing $\hbar \omega_A$.  The response is peaked at the resonance frequency 
$\omega_r \equiv \pi/T$ and has a bandwidth $\sim\!\omega_r/Q$.  The peak phase shift on resonance ($\omega=\omega_r$) has amplitude
\begin{equation}
\Delta\phi_\text{res}=2Q k_\text{eff} h L \sinc\!{\big(\tfrac{\omega_r n L}{2 c}\big)} \cos\!{\big(\tfrac{\omega_r (n-1) L}{2 c}\big)}\label{Eq:PeakPhaseShift}
\end{equation}
which in the low frequency limit $\omega_r \ll \tfrac{c}{n L}$ reduces to $\Delta\phi_\text{res}\approx 2 Q k_\text{eff} h L$.  The phase response shows an $n$-fold sensitivity enhancement from LMT and a $Q$-fold enhancement from operating in resonant mode.  The interferometer can be switched from broadband to resonant mode by changing the pulse sequence used to operate the device (changing $Q$).

To generate the sensitivity curve from the phase response, we find the minimum strain $h$ that can be resolved in the presence of a given phase noise amplitude spectral density $\overline{\delta\phi}_a$.  At each frequency we optimize the LMT enhancement $n$ and resonant enhancement $Q$ while obeying the detector design constraints.  These constraints include limiting the total number $n_p = 2Q(2n-1)+1$ of pulses, the atom wavepacket separation, photon shot noise, and the interferometer duration (see Table~\ref{Tab:designReqs}) \cite{graham2016resonant}.

Figure~\ref{fig:sensitivity} shows the strain sensitivity for two example resonant curves (thick black, narrow) as well as the broadband sensitivity (thick black, wide).  The resonant curve at $f=0.03~\text{Hz}$ uses $8\hbar k$ atom optics with resonant enhancement $Q=9$, while the curve shown at $f=1~\text{Hz}$ uses $1\hbar k$ and $Q=300$.  The broadband curve uses interrogation time $T=7.5~\text{s}$ with $20\hbar k$ atom optics and $Q=1$, so it requires substantially fewer laser pulses.  The brown line at the bottom of the shaded region indicates the envelope of possible resonant response curves that respect the detector design constraints. Since the pulse sequences are defined in software, the detector resonant frequency can quickly be changed to reach any operating point along this line, and the detector can be rapidly switched between broadband and resonant mode.

\subsection{Instrument error model}\label{Sec:ErrorModel}

The instrument error model is summarized below and in Table~\ref{Tab:designReqs}.

\begin{figure}[h]
\centering
		\includegraphics[width=0.6\textwidth]{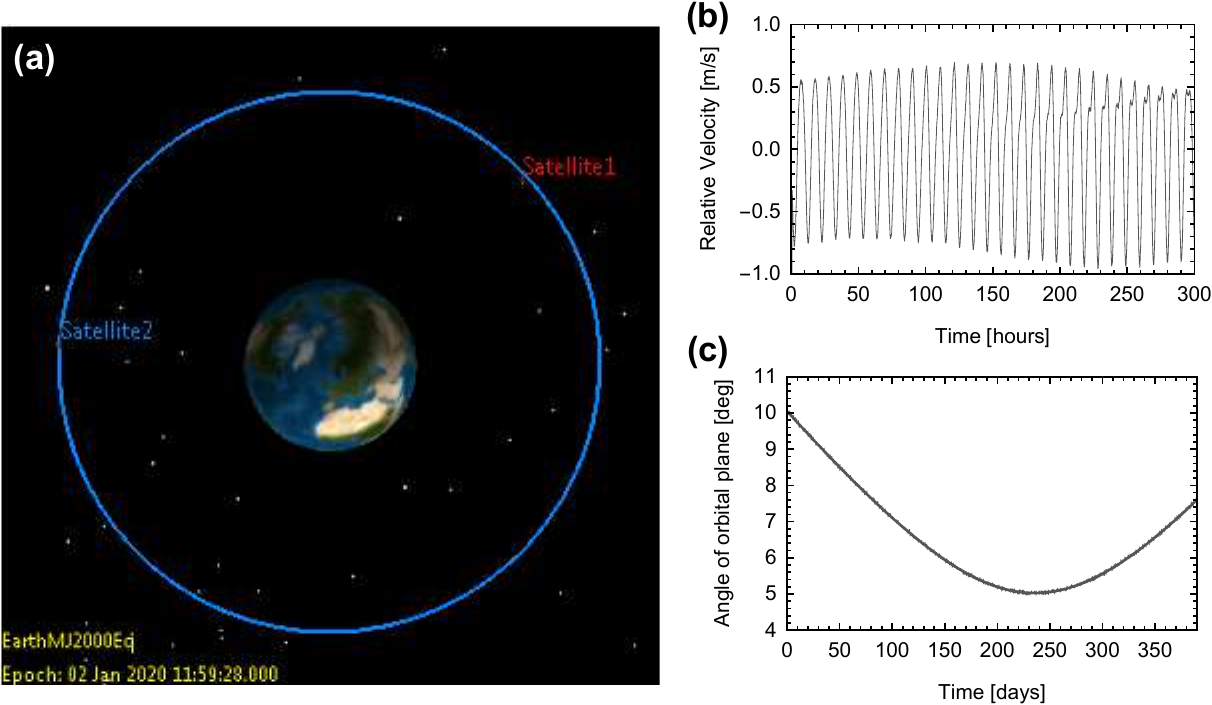}
		\caption{
			Mission orbit analysis. 
            (a) GMAT simulation output showing the circular, geocentric orbits of the two spacecraft.
			(b) Relative velocity of the spacecraft along the direction of the baseline.
			(c) Angle of the orbital plane of the spacecraft with respect to the ecliptic over one year.  Variation of the orbital plane is beneficial for avoiding detector blind spots.
		}
		\label{Fig:scienceOrbit}
\end{figure}

{\vspace{2 pt} \noindent \bf  Vibration Noise} The beamsplitter is rigidly connected to the satellite bus, so any platform vibration noise will affect the beamsplitter as well.  The heterodyne reference ensures that the detector is insensitive to the direct phase noise introduced by vibration of the beamsplitters \cite{hogan_atom_2015}. However, laser wavefront aberrations \cite{Hogan2011, Dimopoulos2008, Bender2011, Dimopoulos2011, graham2016resonant} can lead to a coupling of vibration noise to the atoms. This limits the acceptable transverse vibration noise (perpendicular to the baseline direction) to the level of
\be\hspace{-40pt}
\overline{\delta a}_\perp = 3\times 10^{-10}~\!\frac{\text{m}/\text{s}^2}{\sqrt{\text{Hz}}} \, \Bigg(\!\frac{250}{n_p}\!\Bigg)\Bigg(\!\frac{\Lambda}{1~\text{cm}}\!\Bigg)\!\Bigg(\!\frac{\lambda/100}{\delta \lambda}\Bigg)\!\left(\!\frac{\overline{\delta\phi}_a}{10^{-4}~\text{rad}/\sqrt{\text{Hz}}}\!\right)\left(\frac{f}{0.03~\text{Hz}}\right)^2.
\ee
Spacecraft motion can also couple to the atoms gravitationally \cite{hogan_atom_2015}.  Limits on the allowed longitudinal acceleration depend on the residual gravity gradient $T_{zz}$ along the baseline direction and are given by
\be
\overline{\delta a}_\parallel = 10^{-9}~\!\frac{\text{m}/\text{s}^2}{\sqrt{\text{Hz}}} \, \Bigg(\!\frac{10^3}{n_p}\Bigg)\!\Bigg(\!\frac{10~\text{E}}{T_{zz}}\Bigg)\!\left(\!\frac{\overline{\delta\phi}_a}{10^{-4}~\text{rad}/\sqrt{\text{Hz}}}\!\right)\left(\frac{f}{0.03~\text{Hz}}\right)^4.
\label{Eq:GGnoise}
\ee
Trim masses can be used to reduce the spacecraft gravity gradient to required levels \cite{hogan_atom_2015}.  Acceleration noise can also couple to the detector if there is a non-zero relative velocity $\Delta v$ between the two interferometers \cite{Graham2013a}, as may arise because of perturbations to the spacecraft orbits (see Fig.~\ref{Fig:scienceOrbit}). This coupling constrains platform acceleration noise to $\overline{\delta a}_\parallel <\, 5\times 10^{-8}~\tfrac{\text{m}/\text{s}^2}{\sqrt{\text{Hz}}}\,\, \Big(\!\tfrac{10^3}{n_p}\!\Big)\!\left(\!\tfrac{1~\text{m/s}}{\Delta v}\!\right)\!\left(\!\tfrac{\overline{\delta\phi}_a}{10^{-4}~\text{rad}/\sqrt{\text{Hz}}}\!\right)\!\left(\!\tfrac{f}{0.03~\text{Hz}}\!\right)^2$, which is less demanding than Eq.~\ref{Eq:GGnoise}.

{\vspace{2 pt} \noindent \bf Photon Shot Noise} The heterodyne phase reference is ultimately limited by photon shot noise of the received reference beam light.  Since both atom interferometers are driven using high intensity local lasers, photon shot noise does not directly contribute to phase noise of the atom interferometer signals.  Rather, photon shot noise limits the ability to measure the heterodyne beat note during the finite duration of each interferometer pulse \cite{hogan_atom_2015}.  To avoid limiting the strain resolution, the photon shot noise contribution to the detector noise budget must be less than atom shot noise $\overline{\delta\phi}_a$.  This requires a telescope diameter of
\be
d=30~\text{cm}\,\Big(\!\tfrac{L}{4.4\times 10^7~\!\text{m}}\!\Big)^{\!\frac{2}{5}}  \Big(\!\tfrac{1~\!\text{W}}{P_t}\!\Big)^{\!\frac{1}{10}}  \Big(\!\tfrac{1~\!\text{Hz}/10^3}{f_R/n_p}\!\Big)^{\!\frac{1}{5}}  \Big(\!\tfrac{10^{-4}/\sqrt{\text{Hz}}}{\overline{\delta\phi}_a}\!\Big)^{\!\frac{2}{5}}
\ee
assuming a transmitted power of $P_t = 1~\text{W}$ and a GW sampling rate of $f_R=1~\text{Hz}$. This corresponds to a received power at the other end of the $L=4.4\times 10^7~\!\text{m}$ baseline of $7~\text{$\mu$W}$.

{\vspace{2 pt} \noindent \bf  Pulse timing jitter} In the heterodyne link, any timing delay $t_d$ between the received reference light and the pulses emitted by the phase-referenced local oscillator laser can lead to imperfect cancellation of laser noise.  Noise can arise from either timing jitter $\overline{\delta t}_d$ or frequency noise $\overline{\delta\omega}$, giving a noise PSD of $\overline{\delta\phi}_\text{delay}^2=(n_p \, t_d \,\overline{\delta\omega})^2 + (n_p \, \Delta \,\overline{\delta t}_d)^2$, where $\Delta = \omega-\omega_a$ is the pulse detuning and $n_p$ is the number of laser pulses. Keeping this below atom shot noise $\overline{\delta\phi}_a$ requires noise amplitude spectral densities of
\begin{eqnarray}
\overline{\delta\omega} =&\, 2\pi\!\times\! 10~\!\frac{\text{Hz}}{\sqrt{\text{Hz}}}\,\, \Bigg(\frac{10^3}{n_p}\Bigg)\!\left(\!\frac{1~\text{ns}}{t_d}\!\right)\!\left(\!\frac{\overline{\delta\phi}_a}{10^{-4}~\text{rad}/\sqrt{\text{Hz}}}\!\right)\\
\overline{\delta t}_d =&\, 1~\!\frac{\text{ns}}{\sqrt{\text{Hz}}}\,\, \Bigg(\frac{10^3}{n_p}\Bigg)\!\left(\frac{10~\text{Hz}}{\Delta/2\pi}\right)\!\left(\!\frac{\overline{\delta\phi}_a}{10^{-4}~\text{rad}/\sqrt{\text{Hz}}}\!\right)
\end{eqnarray}

{\vspace{2 pt} \noindent \bf  Pointing jitter} Satellite and laser pointing jitter introduce noise proportional to the offset $\Delta x$ of the atomic ensemble from the baseline axis \cite{Hogan2011}.  The laser pointing requirement is
\be
\overline{\delta\theta} = 10~\!\frac{\text{nrad}}{\sqrt{\text{Hz}}} \, \Bigg(\!\frac{10^3}{n_p}\Bigg)\!\Bigg(\!\frac{1~\text{$\mu$m}}{\Delta x}\!\Bigg)\!\left(\!\frac{\overline{\delta\phi}_a}{10^{-4}~\text{rad}/\sqrt{\text{Hz}}}\!\right).
\label{Eq:pointingReq}
\ee
A feedback loop consisting of the tip-tilt mirror and wavefront sensor in each spacecraft is responsible for maintaining this level of pointing stability in the presence of anticipated spacecraft rotation noise.  The performance of the angle control loop is ultimately limited by the shot noise of the received reference wavefront. The photon shot noise limit for a measurement of the beam angle is $\overline{\delta(\Delta\theta)} = 0.05~\!\frac{\text{nrad}}{\sqrt{\text{Hz}}}\,\left(\frac{30~\text{cm}}{d}\right)\!\left(\!\frac{\overline{\delta\phi}_a}{10^{-4}~\text{rad}/\sqrt{\text{Hz}}}\!\right)$ for a $d=30~\text{cm}$ telescope, assuming (worst case) that the photon shot noise is equal to the atom shot noise $\overline{\delta\phi}_a$. This is sufficient for the required angle tolerance. Pointing requirements for the satellite bus itself are reduced compared to Eq.~\ref{Eq:pointingReq}, requiring pointing stability $\sim 10^{-6}~\text{rad}/\sqrt{\text{Hz}}$, limited by the dynamic range of the tip-tilt mirror servos.

{\vspace{2 pt} \noindent \bf  Temperature stability} Tight ($\sim$ 0.1 mK/Hz$^{1/2}$) thermal control over the temperature of the beam splitter is required to suppress thermally induced path length fluctuations of the glass. These arise from thermal expansion of the glass and temperature induced changes in the index of refraction of the glass.

{\vspace{2 pt} \noindent \bf  Other environmental backgrounds} Blackbody radiation can cause a frequency shift of the atomic transition.  This can result in phase noise in the interferometer if the temperature of the spacecraft fluctuates. For the strontium clock transition, the blackbody shift has a temperature coefficient of $-2.3~\text{Hz}~(T/300 \text{K})^4$ \cite{Falke2011}.  For example, at $T = 30~\text{K}$, this implies a temperature stability requirement of $\lesssim 3~\text{mK}/\sqrt{\text{Hz}}$ for the target strain sensitivity.  Magnetic fields in the interferometry region cause Zeeman shifts of the atomic energy levels, so magnetic field fluctuations are a source of noise.  Simultaneous or interleaved interrogation of each of the linear Zeeman sensitive transitions, as described in Ref. \cite{Falke2011}, results in a  residual quadratic Zeeman coefficient of $-0.23~\text{Hz}/\text{G}^2$ \cite{Falke2011} and also enables measurement of the residual magnetic field.  This suggests that a residual field at the milliguass level with $\sim\text{mG}/\sqrt{\text{Hz}}$ noise is sufficient for the target strain.

\subsection{Clock-based proposals}

In the above proposed architecture, atoms serve a dual role both as an inertial reference and as an optical frequency reference.  The recent successful demonstration of very high performance drag-free satellite control in the LISA Pathfinder Mission \cite{PhysRevLett.116.231101} suggests a variant atom-based architecture in which atoms are used to implement the optical frequency reference, while the inertial reference is implemented using conventional drag free technology.  Such a configuration also enables two-satellite detectors, and has a resonant operating mode as described above.  This approach is described in Refs \cite{kolkowitz2016gravitational,norcia2017new}.  

\subsection{Orbit}

It appears that geocentric orbits are well suited to the above science goals.  As example orbits, consider two spacecraft in geocentric orbit, as shown in Fig.~\ref{Fig:scienceOrbit}(a).  They follow circular orbits with semi-major axis $2.4\times 10^7~\text{m}$.  Satellite 1 has $0~\text{degrees}$ true anomaly and satellite 2 has $138~\text{degrees}$ true anomaly.  Both satellites have inclination $28.5~\text{degrees}$. The baseline is along the line connecting the two spacecraft (line not shown) and has length $L=4.4\times 10^7~\text{meters}$.  The orbits are chosen so that the laser path along the baseline is sufficiently far from of the Earth's upper atmosphere.

The period of the orbits is $600~\text{minutes}$.  The baseline reorients by 90 degrees in one quarter of this time, $150~\text{minutes}$.  For a single baseline detector, this rapid reorientation of the baseline is useful for determining the sky position and polarization of sources.

The nominally circular orbits are perturbed by, for example, tidal forces from the Moon and Sun. This was quantified by simulating the orbits using the NASA General Mission Analysis Tool (GMAT) software package.  The simulation output in Fig.~\ref{Fig:scienceOrbit}(b) indicates that the maximum relative velocity of the satellites (along the baseline direction) is $<1~\text{m/s}$. Spacecraft acceleration noise coupling due to these velocity variations does not appear to limit the detector sensitivity (Section~\ref{Sec:ErrorModel}).

\begin{table}[!htbp]
\scriptsize
\centering
{  \def\arraystretch{1.4}
\begin{tabular}{|c|>{\raggedright\arraybackslash}m{3.7cm}|c|m{7.1cm}|}
	\hline  & \textbf{Parameter} &  \textbf{Constraint} & \textbf{Notes} \\

	\hline  \parbox[t]{3mm}{\multirow{3}{*}{\rotatebox[origin=c]{90}{Spacecraft}}} &  Satellite longitudinal acceleration noise & $10^{-9}~\frac{\text{m}/\text{s}^2}{\sqrt{\text{Hz}}}\left(\frac{f}{0.03~\text{Hz}}\right)^4$ & Residual gravity gradient $T_{zz}<10~\text{E}$\\ 
	\cline{2-4} &  Satellite transverse acceleration noise & $3\times 10^{-10}~\frac{\text{m}/\text{s}^2}{\sqrt{\text{Hz}}}\left(\frac{f}{0.03~\text{Hz}}\right)^2$ & $\lambda/100$ wavefront at $\Lambda=1~\text{cm}$\\ 
	\cline{2-4} &  Satellite pointing stability &  $1~\text{$\mu$rad}/\sqrt{\text{Hz}}$ & \\ 
	\hline

	\hline  \parbox[t]{3mm}{\multirow{4}{*}{\rotatebox[origin=c]{90}{Payload}}} & Telescope aperture &  $30~\text{cm}$ & $\lambda/100$ wavefront aberration\\ 
	\cline{2-4}  & Laser pointing stability &  $10~\text{nrad}/\sqrt{\text{Hz}}$ & Tip-tilt mirror servo; atom positioning $\Delta x < 1~\text{$\mu$m}$\\ 
	\cline{2-4}  & Laser power &  $2~\text{W}$ & Primary interferometry laser; one per spacecraft\\ 
	\cline{2-4}  & Interrogation region length & $1~\text{m}$ & Length of UHV vacuum chamber\\ 

	\hline
    
    \hline  \parbox[t]{3mm}{\multirow{1}{*}{\rotatebox[origin=c]{90}{Atom Interferometry}}}  & Maximum interferometer duration &  $2TQ < 300~\text{s}$ & Limited by vacuum, spontaneous emission\\ 
	\cline{2-4}  & Maximum atom optics pulses &  $n_p < 10^3$ & Includes all LMT and resonant enhancement\\ 
	\cline{2-4}  & Maximum wavepacket separation &  $\frac{n\hbar k}{m}T < 1~\text{m}$ & \\ 
	\cline{2-4}  & AI readout noise & $10^{-4}~\text{rad}/\sqrt{\text{Hz}}$ & Requires flux of $10^8~\text{atoms/sec}$\\ 
	
	\hline 
    
\end{tabular}
}
\caption{Design requirements. All specifications are in the frequency range $0.03~\text{Hz}$ -- $3~\text{Hz}$.  Frequency scalings are indicated where appropriate.}\label{Tab:designReqs}
\end{table}

\section{Discussion}
The above discussion indicates the plausibility of a scientifically significant mid-frequency band gravitational wave detector based on precision atom sensors.  Future work is required to validate core atomic physics subsystems and to fully assess the realism of the proposed configuration.  For example, core atomic physics concepts underlying the proposed atom interferometry and atomic clocks methods have been recently demonstrated.  These include zero-dead time, interleaved, clock operation \cite{schioppo2016ultra}, large momentum transfer atom interferometry \cite{kovachy2015quantum}, preparation of ultra-cold ensembles with delta-kick cooling \cite{Kovachy2014}, phase shear detection for atomic phase read-out \cite{sugarbaker_enhanced_2013}, dynamically decoupled/resonant pulse sequences \cite{Ye_dynamic_decoupling}, high brightness atomic sources \cite{stellmer2013laser}, and single-photon atom interferometry using the strontium clock transition \cite{hu2017atom}.  Still required are demonstrations that combine these methods for gravitational wave specific applications and detailed verification of the instrument error model.  Finally, while the science opportunities in the mid-band described here appear substantial, detailed mission sensitivity and observation strategy studies are still needed.

\ack

We would like to acknowledge collaborators
Adam Black, Teddy Cheung, Lee Feinberg,  Leo Hollberg, Matthew Kerr, William Klipstein, Tom Loftus, Michael Lovellete, Mikael Lukin, Peter Michelson, Babak Saif, Bernard Seery, Alex Sugarbaker, Ronald Walsworth, Kent Wood, and Jun Ye.

SR was supported in part by the NSF under grants PHY-1638509 and PHY-1507160, the Alfred P. Sloan Foundation grant FG-2016-6193 and the Simons Foundation Award 378243. PWG acknowledges the support of NSF grant PHY-1720397 and DOE Early Career Award DE-SC0012012.  This work was supported in part by Heising-Simons Foundation grants  2015-037 and  2015-038.  MK and PWG acknowledge support from the Keck Foundation.  MK acknowledges support from the NASA/JPL Consortium for Ultracold Atoms in Space (JPL \#1548507).  RWR was supported in part by NASA Grant 80NSSC17K0024.

\section*{References}

\bibliographystyle{iopart-num}
\bibliography{GW_reference}

\end{document}